\documentclass{qnp2005}

\newcommand{\ket}{\rangle}
\newcommand{\bra}{\langle} 
\newcommand{\beq}{\begin{eqnarray}}
\newcommand{\eeq}{\end{eqnarray}}
\newcommand{\del}{\partial}
\newcommand{\Slash}[1]{\ooalign{\hfil/\hfil\crcr$#1$}}

\begin{document}
\setcounter{page}{1}

\title{Path-integral approach in a chiral quark-diquark model to the
nucleon structure and interactions}

\author{Keitaro Nagata and Atsushi Hosaka}

\address{Research Center for Nuclear Physics (RCNP), Osaka University\\
Ibaraki, Osaka 567-0047, Japan
\\ 
E-mail: nagata@rcnp.osaka-u.ac.jp}

\maketitle

\abstracts{
We study the structure of the nuclear force by using a path-integral
hadronization approach in a chiral quark-diquark model. 
After the construction of the chiral quark-diquark model, we hadronize it
to obtain a meson-baryon Lagrangian. The effective meson-baryon Lagrangian
incorporates chiral symmetry and the composite description of
the mesons and baryons. Using the effective meson-baryon Lagrangian we
investigate the structure of the nuclear force in the simple case of
neglecting the axial-vector diquark. It is shown that the
meson-baryon Lagrangian contains two kinds of the nuclear force; the
meson-exchange interaction and a quark-diquark loop interaction. It is
also shown that the quark-diquark loop interaction consists of the
scalar and vector interactions. The properties of these interactions
are discussed.
}

\section{Introduction}

The origin of the short range repulsion in the nucleon-nucleon($NN$)
interaction is a long-standing problem.
The $NN$ interaction below the meson production threshold 
is phenomenologically well understood, having been fit to the phase shift 
analysis, but its microscopic understanding is still needed.  
While the long range part is well described by the meson exchange 
picture, there are several different approaches to the description of 
the short range part, including meson exchange and quark 
exchange~\cite{Lacombe:dr,Machleidt:hj,Toki:ai,Oka:rj,Takeuchi:yz,Fujiwara:2001pw}.
Because in the short range region the two nucleons have a substantial
spatial overlap, it is commonly assumed that the internal structure of the
nucleon gives a sizable contribution to the nuclear force at such
short distances. To obtain a comprehensive understanding of the
nuclear force from the short to long range distances, models which 
incorporate both the meson exchange and the internal struture of 
the nucleon are needed.

In a recent publication, we studied the structure of the nuclear force
using the path-integral hadronization approach to a chiral
quark-diquark model~\cite{Nagata:2003gg}. This method incorporates 
two aspects of chiral symmetry, which naturally describes the pion
exchange interaction, and of the internal structure of hadrons. 
This method employs an extended model of the 
Nambu-Jona-Lasinio type~\cite{NJL} with the interactions that are 
not only of the quark-antiquark type
but also of the quark-diquark type~\cite{Abu-Raddad:2002pw}.  
The quarks and diquarks were integrated out to generate 
an effective Lagrangian for mesons and baryons while maintaining 
important symmetries, such as the gauge and chiral symmetries.  
The hadron structure was then described in terms of its constituents:
a quark and an antiquark for mesons and a quark and a diquark for baryons.  
It was also pointed out that the resulting effective Lagrangian 
contains various interactions among hadrons, such as meson-meson, 
meson-baryon and baryon-baryon interactions.  

In our model, all components of the $NN$ force are contained in 
an effective Lagrangian that is written in a concise form
as a trace-log.  
Then, the expansion of the trace-log terms produces an NN force that is
described as meson exchanges at long and medium ranges and
quark-diquark exchanges at short ranges.
The latter was then shown to contain various types of non-local
interactions, including a scalar iso-scalar type and a vector
iso-scalar type.  We evaluated the properties of the scalar and vector
type interactions; their ranges, effective masses and strengths.

In this paper, we report the results of the study of the nuclear
force in the simple framework where the axial-vector diquark is neglected. 
In the previous work, we suggested that the neglecting the
axial-vector diquark does not affect to the size of the nucleon, hence
we can properly evaluate the ranges of 
the $NN$ interactions with only scalar diquarks.

We organize this paper as follows.  
In $\S$ 2, we briefly give the derivation of the trace-log 
formula in the path-integral hadronization of the NJL model
with quark-diquark correlations.  
In $\S$ 3, terms containing the $NN$ interaction are investigated 
in detail, 
where the general structure of the $NN$ amplitude is presented.  
We present a sample numerical calculation for the case 
in which there is only a scalar diquark.  
The present study of the $NN$ interaction is not quantitatively 
complete, but it will be 
useful in demonstrating some important aspects of the nuclear force, 
in particular that the range of the short range interaction is 
related to the intrinsic size of the nucleon.  
The final section is devoted to a summary.

\section{Effective Lagrangian for mesons and nucleons}

We now briefly review the method to derive 
an effective Lagrangian for mesons and nucleons  
from a quark and diquark model of chiral symmetry following the 
previous work of Abu-Raddad et al~\cite{Abu-Raddad:2002pw}.
We start from the NJL Lagrangian,
\begin{equation}
{\cal L_{\mbox{NJL}}} = {\bar q} (i\rlap/\partial -m_0) q + \frac{G}{2} \left[
({\bar q} q)^2 + ({\bar q} i \gamma_5 \vec{\tau} q)^2 \right]\;.
\end{equation}
Here, $q$ is the current quark field, $\vec{\tau}$ represents the isospin
(flavor) Pauli matrices, $G$ is
a dimensional coupling constant, and $m_0$ is the  
current quark mass. 
In this paper, we set $m_0 = 0$ i.e. we work in the chiral limit.  
As usual, the NJL Lagrangian is bosonized 
by introducing meson fields as collective auxiliary fields in the 
path-integral method~\cite{Eguchi:1976iz,Dhar:1983fr,Ebert:1986kz}.  
At an intermediate step, we find the following Lagrangian:  
\beq
{{\cal L}^\prime_{q\sigma\pi}} = 
\bar{q}\left( i\rlap/\partial  
- (\sigma + i\gamma_5 \vec\tau\cdot\vec\pi) \right) q
-\frac{1}{2G}(\sigma^2+\vec\pi^{\; 2})\, . 
\label{Lprime}
\eeq
Here $\sigma$ and $\vec \pi$ are 
scalar-isoscalar sigma
and pseudoscalar-isovector pion fields, as generated from 
$\sigma \sim \bar q q$ and $\vec \pi \sim i\bar q \vec \tau \gamma_5q$, 
respectively.  
For our purpose, 
it is convenient to work in a non-linear basis rather than the
linear one~\cite{Ebert:1997hr,Ishii:2000zy}.
This is realized through the chiral rotation from the current ($q$) to 
constituent ($\chi$) quark fields: 
\beq
\chi = \xi_5 q\, , \; \; \; \; 
\xi_5 
= 
\left(
\frac{\sigma + i \gamma_5 \vec \tau \cdot \vec \pi}{f} 
\right)^{1/2} \, , 
\eeq
where $f^2 = \sigma^2 + \vec \pi^{\; 2}$.  
Thus, we find 
\beq
{\cal L^\prime_{\chi \sigma \pi}} = 
\bar{\chi}\left(i\rlap/\partial  
- f - \rlap/v - \rlap/a\gamma_5\right) \chi
-\frac{1}{2G} f^2 \, , 
\label{Lprime2}
\eeq
where 
\beq
v_\mu = - \frac{i}{2} \left(
\del_\mu \xi^\dagger \xi +  \del_\mu \xi \xi^\dagger 
\right)\, , \; \; \; 
a_\mu = - \frac{i}{2} \left(
\del_\mu \xi^\dagger \xi -  \del_\mu \xi \xi^\dagger 
\right)\,  
\eeq
are the vector and axial-vector currents written in terms of the 
chiral field,
\beq
\xi 
= 
\left(
\frac{\sigma + i \vec \tau \cdot \vec \pi}{f} 
\right)^{1/2}\, .
\eeq
The Lagrangian (\ref{Lprime2}) describes not only the 
kinetic term of the quark, but also quark-meson interactions 
such as the Yukawa one, Weinberg-Tomozawa one etc.  

In the model we consider here, we introduce 
diquarks and their interaction terms with quarks.  
We assume local interactions between quark-diquark
pairs to generate the nucleon field.  
As suggested by a method of constructing a local nucleon 
field, it is sufficient to consider two types of diquarks,
a scalar, isoscalar diquark, $D$,  and an 
axial-vector, isovector diquark, $\vec D_\mu$~\cite{espriu}. In this
work, we consider only the scalar diquark.
Hence, our microscopic Lagrangian for quarks, diquarks and 
mesons is given by 
\begin{eqnarray}
{\cal L} = \bar{\chi}(i\rlap/\del - f 
- \rlap/v - \rlap/a \gamma_5) \chi \;-\;
\frac{1}{2G}f^2\;+
D^\dag (\del^2 + M_S^2)D 
+ \tilde{G}\;\bar{\chi} D^\dag D \chi\; .
\label{lsemibos1}
\end{eqnarray}
In the last term, $\tilde G$ is a coupling constant for the 
quark-diquark interaction.   

\begin{figure}
       \centerline{\includegraphics[width=5cm]
                                   {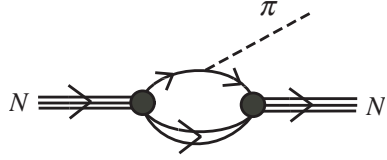}}
\centering
\begin{minipage}{10cm}
   \caption{\small 
   A quark-diquark one-loop diagram for the meson-nucleon Yukawa
   vertex. The solid, double, dotted and triple lines represent the quark,
   diquark, meson and nucleon, respectively. The blobs represent the three point
   quark-diquark-baryon interaction. }
   \label{mbb}
 \end{minipage}
\end{figure}

\begin{figure}
       \centerline{\includegraphics[width=8cm]
                                   {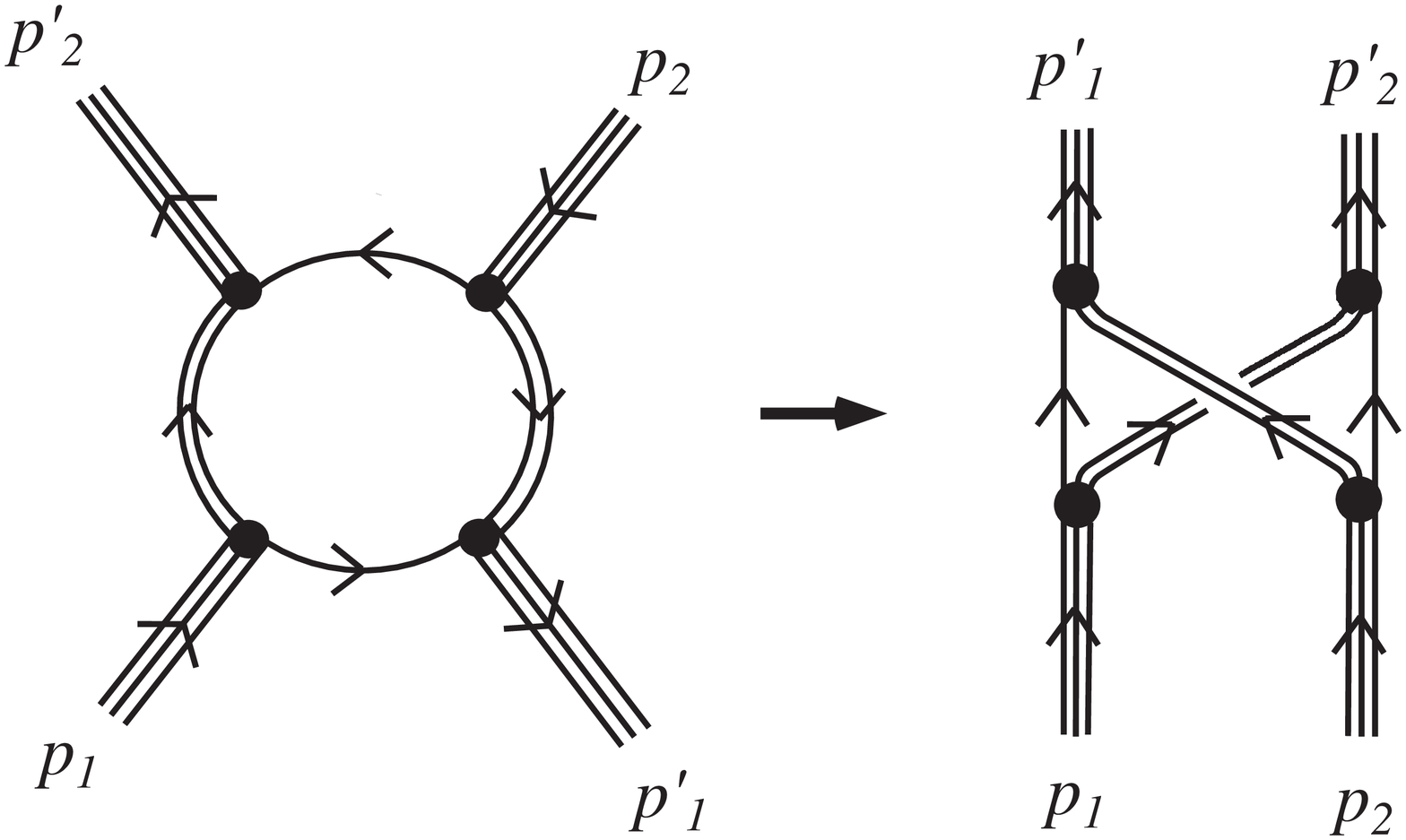}}
\centering
\begin{minipage}{10cm}
   \caption{\small 
   A loop diagram for the $NN$ interaction (left) and the equivalent 
diquark exchange diagram.  }
   \label{bbbbloop}
 \end{minipage}
\end{figure}

Now, the hadronization procedure can be carried out straightforwardly by 
introducing the baryon fields as auxiliary fields, 
$B\sim D\chi$, and by eliminating the 
quark and diquark fields in (\ref{lsemibos1}).  
The final result is written in a compact form as~\cite{Abu-Raddad:2002pw}
\begin{eqnarray}
{\cal L}_{\rm eff} &=& 
- \frac{1}{2G}f^2 
- i \;{\rm tr\; ln} (i \rlap/\del - f -\rlap/v - \rlap/a \gamma_5)
\;-\; \frac{1}{\tilde{G}}\; \bar{B} B \;+\; 
i\;{\rm tr \; ln} ( 1- {\cal S} ) \, .
\label{effL}
\end{eqnarray}
Here, the trace is taken over space-time, color, flavor and Lorentz indices, and 
the operator ${\cal S}$ is defined by 
\begin{eqnarray}
{\cal S} =  \Delta^T\bar{B}\;
  \;S\;B\;,
\label{eqbox}
\end{eqnarray}
In this equation, $S=(i \rlap/\del-f-\rlap/v-\rlap/a\gamma_5)^{-1},\;
\Delta=(\del^2-M_S^2)^{-1}$
are the propagators of the quark and scalar diquark, respectively, and  
transposed diquark propagators, as denoted by 
the superscript $T$, are employed. Note that the quark propagator $S$
contains the interactions with pions. Through this interactions, the
nucleon-pion interactions are obtained. 
Though the effective meson-nucleon Lagrangian (\ref{effL}) 
looks simple, 
it contains many important physical ingredients when the trace-log
terms are expanded:
\begin{itemize}
\item
It generates a meson Lagrangian in a chirally symmetric manner.  
Up to fourth order in the meson fields, it produces precisely 
the Lagrangian of the linear sigma model with the realization of  the 
spontaneous breaking of chiral symmetry. Hence, the vacuum expectation
value $f$ turns out to be the pion decay constant $f_\pi$.

\item
From the second trace-log term, a nucleon effective Lagrangian 
is derived.  
In a previous paper, the kinetic term of the nucleon was 
investigated, and the mass of the nucleon was computed at the one-loop 
level~\cite{Abu-Raddad:2002pw}.

\item
In the nucleon effective Lagrangian, meson-nucleon couplings 
appear through the diagrams, as shown in Fig.~\ref{mbb}.  
Their strengths and form factors can be computed 
with the use of the underlying quark-diquark dynamics.  
Using these vertices, meson-exchange interactions 
are constructed.  the photon-nucleon couplings are also obtained in
the same manner. The radii and magnetic moments were evaluated without
the axial-vector diquark\cite{Abu-Raddad:2002pw,Nagata:2003gg}.

\item
There are diagrams that contain many nucleon fields.  
For instance, $NN$ interactions are expressed 
as one-loop diagrams, as shown in Fig.~\ref{bbbbloop}.  
This term describes the short range part of the $NN$ 
interaction.  
In this paper, we focus our attention mostly on the $NN$ 
interaction derived from the one-loop diagrams.  

\end{itemize}

\section{Short range interaction}
\label{sec:qdloop}

We evaluate the short range interaction described by the 
quark-diquark loop, as shown in Fig.~\ref{bbbbloop}.   
Using the interaction vertices given in Eq.~(\ref{eqbox}), 
it is straightforward to compute the amplitude for the 
quark-diquark loop:
\begin{eqnarray}
&&{\cal M}_{NN} = -i N_c Z^2 
\frac{d^4k}{(2\pi)^4}\nonumber\\
&\times&  \frac{\bar{B}(p_1^\prime)(\Slash{k}+m_q)B(p_1)
\bar{B}(p_2^\prime)(\Slash{p}_2-\Slash{p}_1^{\prime}+\Slash{k}+m_q)B(p_2)}
{[(p_1-k)^2-M_S^2][k^2-m_q^2]
[(p_1^\prime -k)^2-M_S^2]
[(p_2-p_1^{\prime}+k)^2-m_q^2]}\, .
\label{eqn:NNN}
\end{eqnarray}  
where $N_c$ is the number of colors and $Z$ is the wave-function
renormalization constant.
In this equation, the momentum variables are assigned such that 
$p_1$ and $p_1^\prime$ ($p_2$ and $p_2^\prime$) are for the pair of 
contracted baryon fields, 
$\bar B(p_1^\prime) \cdots B(p_1)$ ($\bar B(p_2^\prime) \cdots B(p_2)$), 
and the momentum transfer is defined by 
$q = p_1^\prime -p_1$.
Note that the momentum $q$ is carried by the diquark pair.  
The amplitude defined in this way can be interpreted as a direct term 
in the local potential approximation.  
When computing physical quantities such as phase shifts, we need to 
include the exchange term that is obtained by interchanging the 
momentum variables $p_1^\prime \leftrightarrow p_2^\prime$.  
Although the one-loop integral (\ref{eqn:NNN}) converges 
when the scalar diquark is included, we keep the counter terms of the 
Pauli-Villars regularization.  
Because our model is a cut-off theory with 
a relatively small cutoff mass, $\Lambda_{PV} = 0.63$ GeV, 
the counter-terms play a significant role.
However, because we include only the scalar diquark, we do 
not attempt to make a comparison at the quantitative level.  
Rather, in the following, we study some basic properties of the amplitude
itself, mostly the interaction ranges extracted from Eq.~(\ref{eqn:NNN}).   

Let us evaluate the integral in the center-of-mass system
for elastic scattering:
\begin{eqnarray*}
p_1&=&(E_{\vec{p}},\ \vec{p}),\ p_2=(E_{\vec{p}},\ -\vec{p}),\\
p_1^{\prime}&=&(E_{\vec{p}^{\; \prime}},\ \vec{p}^{\; \prime}),\ 
p_2^{\prime}=(E_{\vec{p}^{\; \prime}},\
-\vec{p}^{\; \prime})\; , \; \; \; 
|\vec p| = |\vec p^{\; \prime}| \; .  
\end{eqnarray*}
To proceed, we write 
the amplitude (\ref{eqn:NNN}) as
\beq
{\cal M}_{NN}
&=&
F_S(\vec{P},\vec{q})(\bar{B}B)^2+F_V(\vec{P},\vec{q})
(\bar{B}\gamma_{\mu}B)^2+\cdots\
\label{eqn:NNforce} ,
\eeq
where
\beq
\label{defPq}
\vec{P}=\vec{p}^{\; \prime}+\vec{p}\, , 
\, \; \; \; 
\vec{q}=\vec{p}^{\; \prime}-\vec{p} \, .
\eeq
Roughly speaking, the $q$-dependence expresses the interaction range in the $t$-channel, 
whose Fourier transform is interpreted as the $r$-dependence of 
the local potential, while 
the $P$-dependence expresses the non-locality of the interaction.

In Eq.~(\ref{eqn:NNforce}), we have defined the coefficients 
$F_S$ and $F_V$ as  
scalar and vector interactions, respectively.  
The ellipsis in Eq.~(\ref{eqn:NNforce}) then 
include terms involving external momenta $p_i$, such as 
$(\bar B \Gamma(p_i) B)(\bar B B)$ and $(\bar B B)(\bar B \Gamma(p_i) B)$, 
where  $\Gamma(p_i)$ are $4 \times 4$ matrices involving $p_i$.  
These momentum dependent terms are, however, expected to play a less important 
role than the dominant components of the scalar and vector terms, as in 
boson exchange models.   
Therefore, we consider here the corresponding terms of scalar and vector 
types.   
It turns out that the interaction coefficients defined in this way
are attractive for $F_S$ and repulsive for $F_V$.

\begin{figure}[tbh]
\centering{
\includegraphics[width=11.0cm]{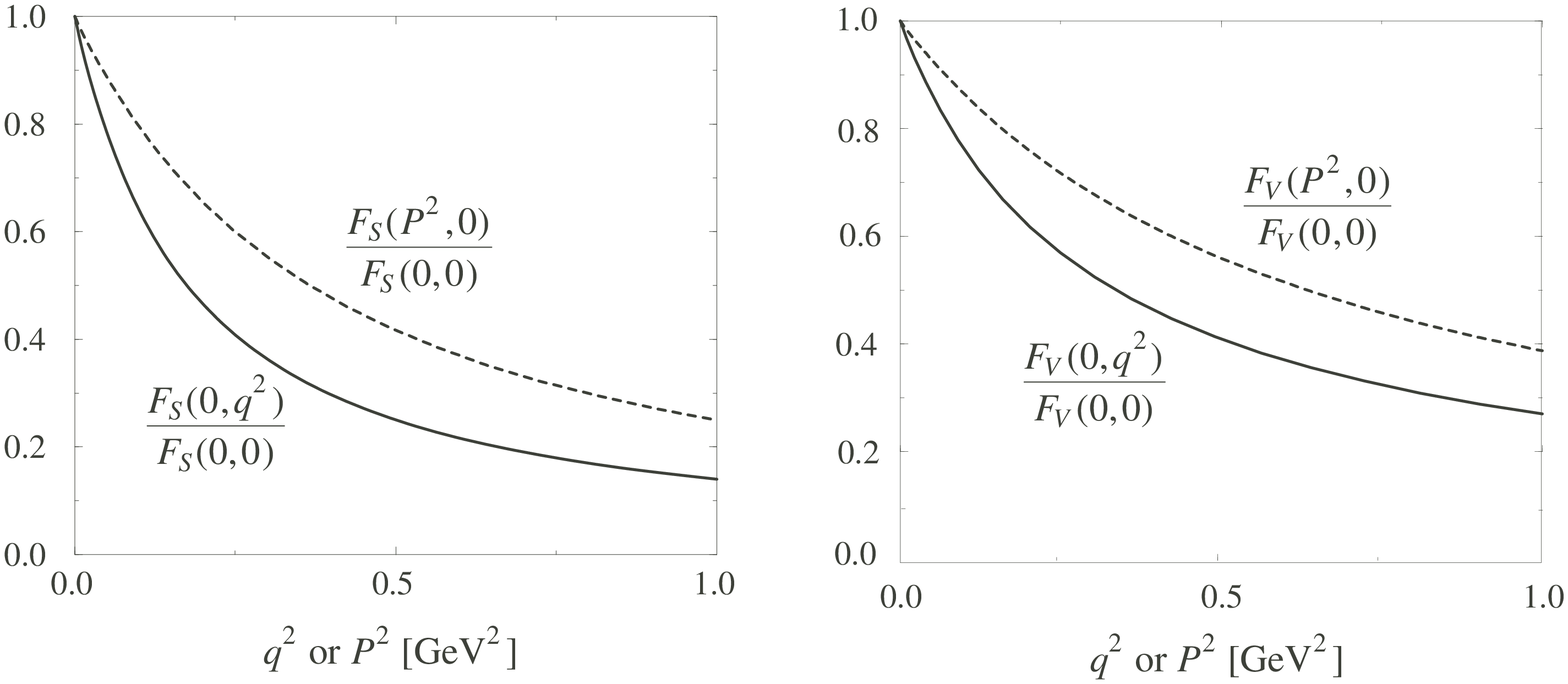}
}
\begin{minipage}{10cm}
\caption{\small 
Normalized scalar and vector form factors
as functions of $q^2 = |\vec q^{\; 2}|$ with fixed 
$\vec P = 0$ (solid curve) and 
those of $P^2 = |\vec P^{\; 2}|$ with fixed 
$\vec q = 0$ (dashed curve). 
The calculations were performed using the parameter set given in
Table 1. }
\label{ffscalar}
\end{minipage}
\end{figure}

As anticipated,  
the amplitude is highly non-local, as the quark-diquark 
loop diagram implies.  
As a matter of fact, the interaction range for the diquark exchange 
in the $t$-channel is shorter than that for 
the quark exchange in the $s$-channel.  
This is shown explicitly in Fig.~\ref{ffscalar}.  
Nevertheless, we proceed further and carry out an expansion in $\vec
P$. We obtain
\beq
\label{Vexpand}
F_{i}(\vec P, \vec q) 
=
F_{i}(\vec P, \vec q)\vert_{\vec P = 0}
+ \vec P \cdot \frac{\del}{\del \vec P} 
F_{i}(\vec P, \vec q)\vert_{\vec P = 0} 
+ \cdots \, ,
\eeq
where $i$ stands for $S$ or $V$.  
The resulting ${\vec q}$ dependent function, in particular the 
first term, can be interpreted as the
Fourier transform of a local potential as a function of the relative 
coordinates ${\vec r} = {\vec x}_1 - {\vec x}_2$.  
Then, the general structure of the 
$NN$ potential can be studied by 
performing the non-relativistic reduction of 
the amplitude, Eqs.~(\ref{eqn:NNforce}) and (\ref{Vexpand}).  
It contains central, spin-orbit and tensor components that accompany
functions of non-locality $\vec{P}$.

We now discuss the  functions 
$F_S$ and $F_V$ in Eq.~(\ref{eqn:NNforce}) for 
the leading-order terms of Eq.~(\ref{Vexpand}).
Because we cannot write the resulting $q$ dependent functions in 
an analytic form,
we have carried out numerical calculations employing 
several different values of $\tilde G$ parameter for different binding 
energies and the size of the quark-diquark bound state.

At this point, we explain our parameters and the regularization scheme. 
We follow the scheme presented in Ref.~\cite{Abu-Raddad:2002pw}, except for the
treatment of $\tilde{G}$. The $\Lambda_{PV}$ is the 
cutoff mass of the Pauli-Villars method used in this work to
regularize divergent integrals.
We note that our model is unrenormalizable.
The NJL coupling constant $G$ and the cut-off mass 
$\Lambda_{PV}$ are fixed to
generate the constituent quark mass $m_q$ and the pion decay constant 
through the NJL gap equation in
the meson sector \cite{Ebert:1986kz,Hatsuda:1994pi,Ebert:1994mf}.
The mass of the scalar diquark $M_S$ is determined in the NJL model
by solving the Bethe-Salpeter
equation in the diquark channel \cite{Vogl:1991qt,Cahill:1987qr}. 
Then, we have one free
parameter, the quark-diquark coupling constant $\tilde{G}$. 
The strength of $\tilde G$ controls the binding or size of 
the nucleon and generates the mass of the nucleon $M_N$.
Then we perform numerical calculations by varying the coupling
constant $\tilde{G}$.
In Table~\ref{tab:parameter}, we list typical parameter values as 
used in Ref.~\cite{Nagata:2003gg}.

\begin{table}[tbp]
\begin{center}
     \caption{Model parameters.}
     {\footnotesize
     \begin{tabular}{ccccccc}
	\hline
  & $M_N$ [GeV] &
	$m_q$ [GeV] & $M_S$ [GeV] & $\Lambda_{PV}$ [GeV]  &
     $\tilde{G}$ [GeV$^{-1}$] &  $\bra r^2 \ket ^{1/2}$[fm]\\
	\hline
	set I & 0.94 & 0.39 & 0.60
	& 0.63 & 271.0 &0.77 \\
        set II & 0.85 &0.39 & 0.60 & 0.63 & 445.9& 0.54\\ \hline
     \end{tabular}
      }\label{tab:parameter}
\end{center}
\end{table}

\begin{figure}[tbh]
\centering{
\includegraphics[width=5cm]{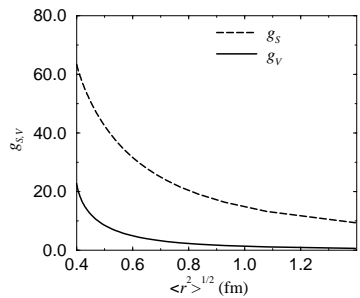}
}
\begin{minipage}{10cm}
\caption{\small 
Scalar and vector coupling constants as functions of the 
nucleon size, $\bra r^2 \ket^{1/2}$. }
\label{gSV}
\end{minipage}
\end{figure}

First, we determine the strengths of the interactions by extracting 
coupling constant squares $g_i^2$ as
\beq
|F_i(0,0)| = \frac{g_i^2}{m_i^2} \, ,\; \; \; \; g_i > 0\, ,
\eeq
where the effective meson masses 
$m_i$ are evaluated using the inverse of (\ref{defrange}) and are plotted 
in Fig.~\ref{fig:sizevsrange}.  
The results are displayed in Fig.~\ref{gSV} as functions of the 
size of the nucleon, $\bra r^2 \ket ^{1/2}$.  
The coupling strengths can be compared with the empirical values
$g_S \sim 10$ and $g_V \sim 13$~\cite{Machleidt:hj}.  
The present results are strongly dependent on $\bra r^2 \ket$.  
When only a scalar diquark is included, the scalar interaction 
becomes much stronger than the vector interaction.  
Phenomenologically, the vector (omega meson) coupling is  
stronger than the scalar (sigma meson) coupling.  

We should comment on the effect of the axial-vector diquark.  
The loop integral containing the axial-vector diquarks diverges 
and has a large numerical value even when it is regularized.  
Therefore, it significantly affects the absolute values of the 
loop integrals as well as that of the baryon self energy, which 
is necessary to extract the normalization factor $Z$.   
Therefore, it is expected that the strengths are affected significantly, but 
the ranges and the corresponding masses, which are computed from 
the momentum dependence of the normalized loop integrals, are 
not affected much.

\begin{figure}[tbh]
\centering{
\includegraphics[width=5cm]{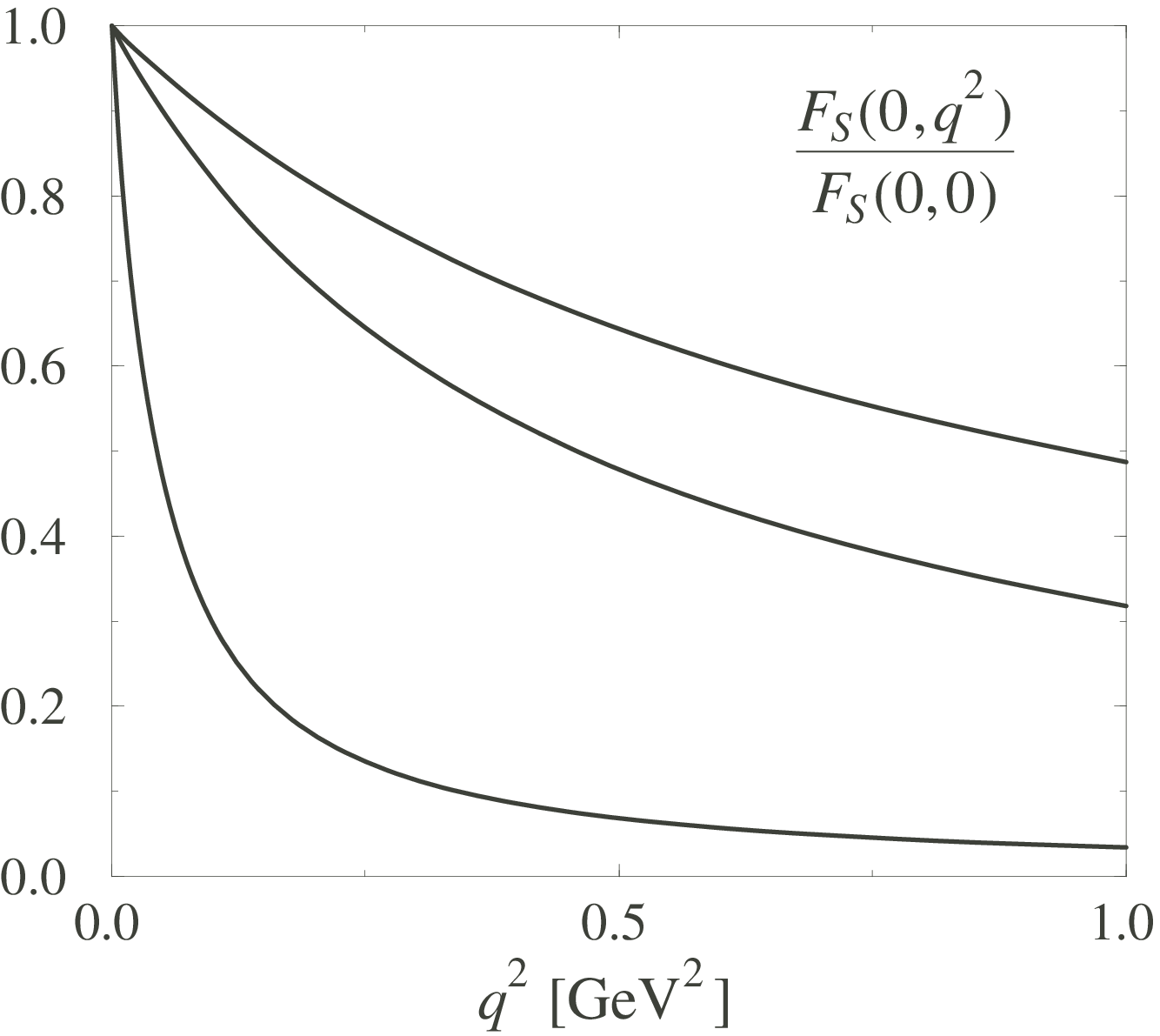}
\quad
\includegraphics[width=5cm]{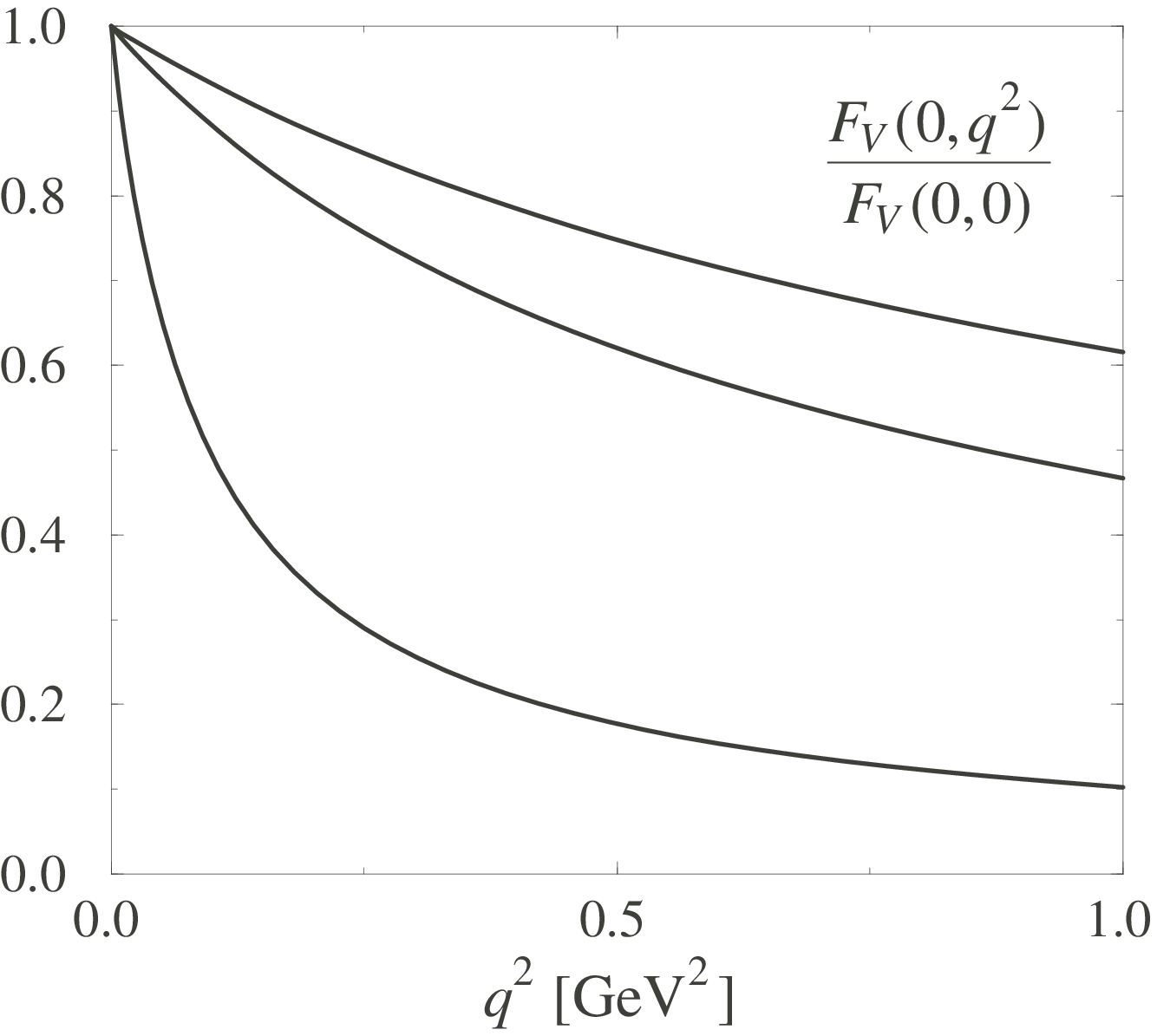}
}
\begin{minipage}{10cm}
\caption{\small 
The scalar (left panel) and vector (right panel) 
form factors $F_{S,V}\ (q^2,\ P^2=0)/F_{S,V}(q^2=0,\
P^2=0)$, (form factors are normalized to 1 at $q^2=0$). 
The curves correspond to $M_N$=0.98, 0.85, 0.70 GeV from bottom 
to top.\label{fig5scalar}}
\end{minipage}
\end{figure}

\begin{figure}
\centering{
\includegraphics[width=5cm]{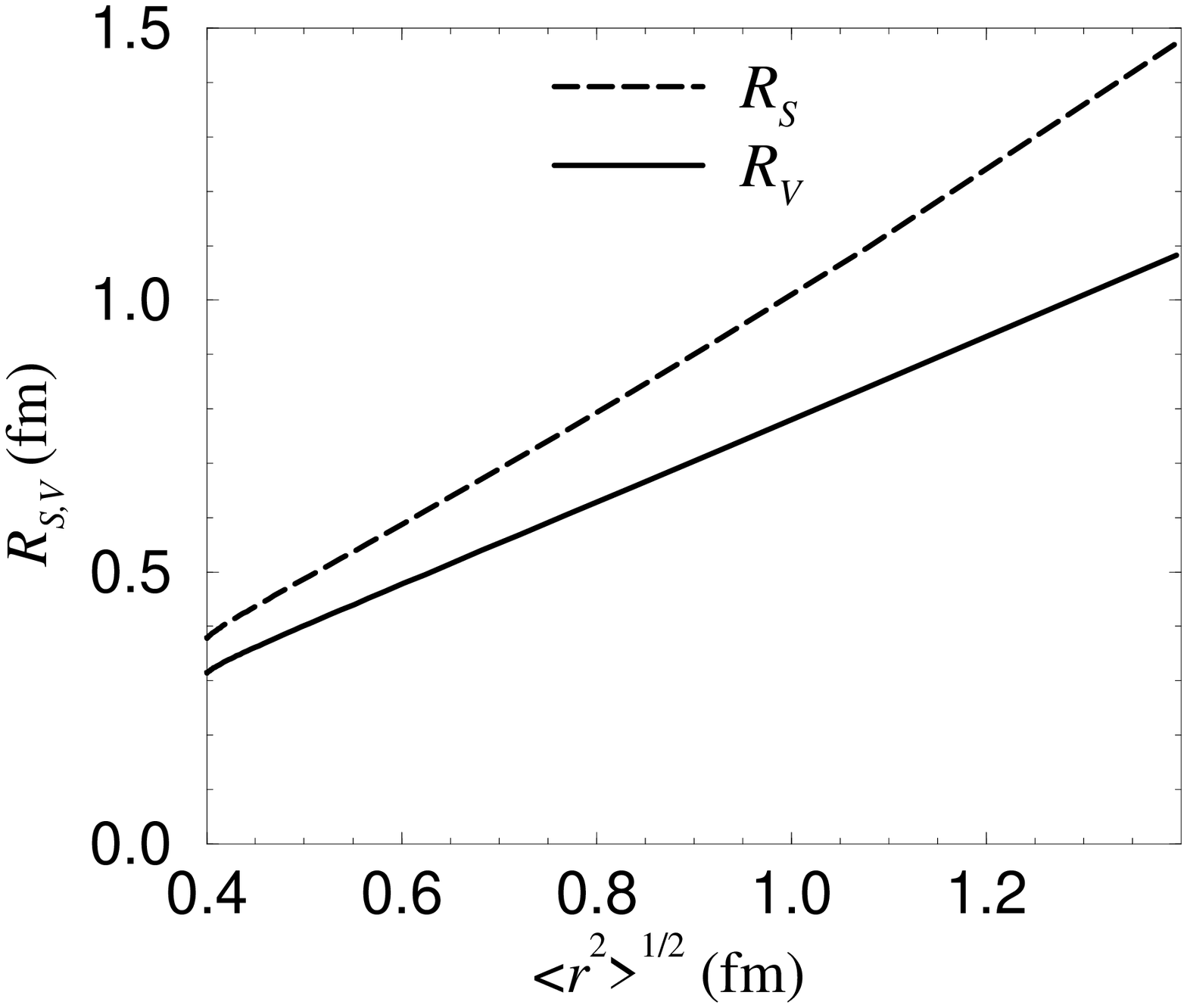}
\hspace*{1cm}
\includegraphics[width=5cm]{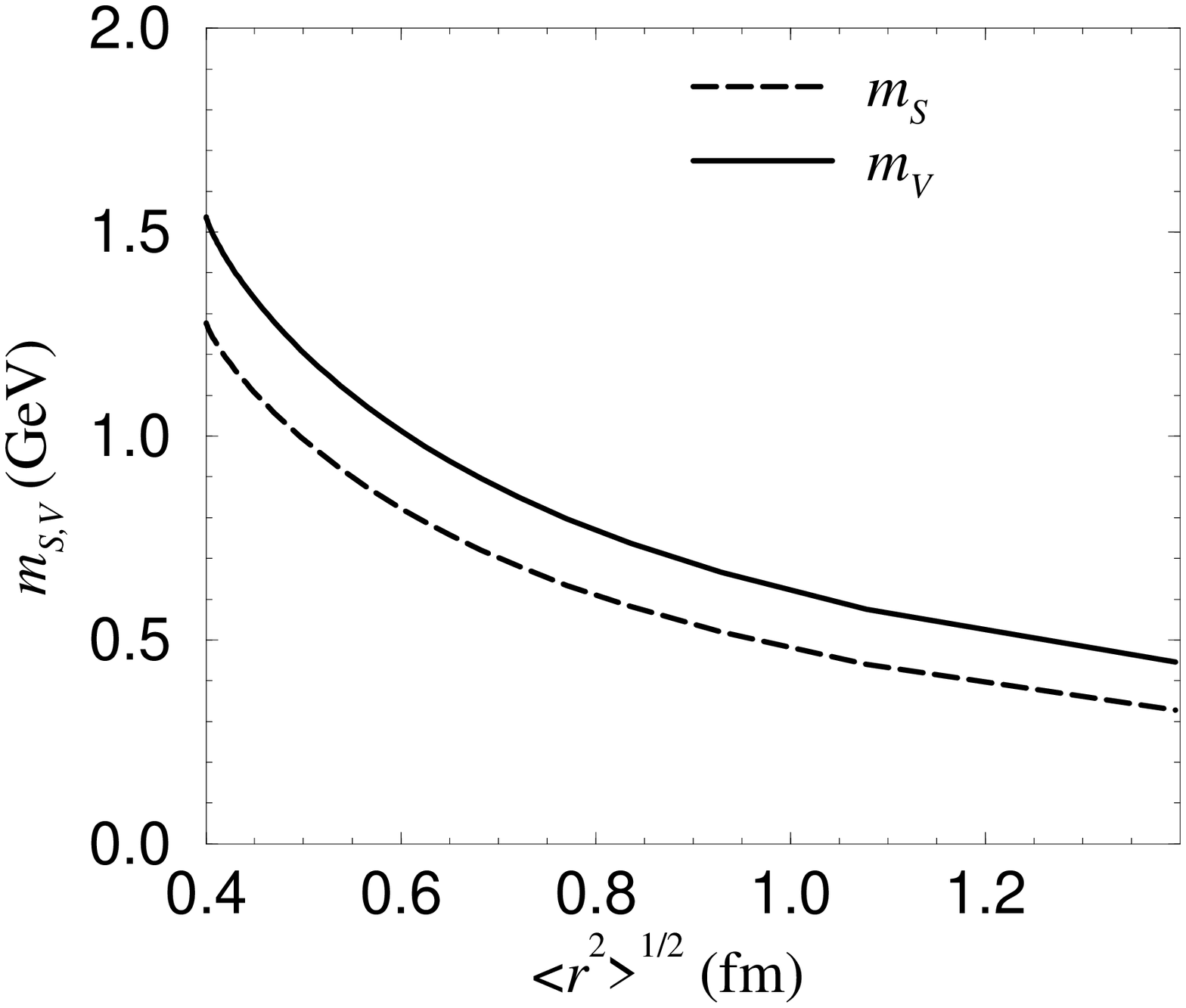}}
\begin{minipage}{10cm}
\caption{\small
 The interaction range $R_i$ (left panel) and the corresponding 
mass parameter $m_i$ (right panel) as functions of the nucleon size
(distribution of a quark and a diquark).}
   \label{fig:sizevsrange}
\end{minipage}
\end{figure}

Let us discuss the interaction ranges.  
In Fig.~\ref{fig5scalar}, we show the $q$-dependence 
of the zeroth order coefficients in Eq.~(\ref{Vexpand}) for three 
different sizes of the nucleon.  
It is obvious from Fig.~\ref{fig5scalar} that
as the size of the nucleon becomes smaller the interaction ranges
become shorter.   
More quantitatively, we define the interaction ranges $R_i$ by 
\beq
R_i^2 \equiv - 6 \left.\frac{1}{F_i(q^2)}\frac{\del F_i}{\del
q^2}\right|_{q^2\rightarrow 0} \, ,  
\label{defrange}
\eeq
which are related to the mass parameters of the interaction ranges as
$m_i \equiv \sqrt{6}/R_i$.  
It is interesting that the range (and hence mass)
parameters of the interactions are approximately 
proportional to the size of the nucleon $\bra r^2\ket^{1/2}$, as shown in Fig.~\ref{fig:sizevsrange}.
When the parameter set I listed in Table~\ref{tab:parameter} is used, 
the mass parameters are about 650 MeV and 800 MeV for the scalar and vector
interactions, respectively, which 
are very close to the masses of the sigma and omega mesons.  
If, however, we use the parameter set II for 
a nucleon size of about 
0.5 fm, then the two masses become 
$m_S \sim 800$ MeV and 
$m_V \sim 1000$ MeV.

In the present analysis, apart from the absolute values of the 
interaction strengths, the interaction ranges for the scalar
and vector interactions have been produced appropriately with the 
quark-diquark loop diagram.  
One question concerns the small but non-negligible difference
between the ranges in the scalar and vector channels, which is
consistent with empirical results.  
This can be roughly understood from the dimensionality of the loop 
integral.  
As seen from Eq.~(\ref{eqn:NNN}), 
the integrand for the vector interaction is of higher 
order with respect to the loop momentum than the scalar interaction.  
Because of this, the vector part reflects shorter distant dynamics 
and produces a shorter interaction range.  

Although we have obtained reasonable results for the interaction
ranges, the short range part of the nuclear force must be repulsive.
Hence, current approach is not satisfactory. We believe that the
inclusion of the axial-vector diquark will give a sizable contribution
to the short range repulsion because the quark-axial-vector
diquark-baryon vertex is basically spin-spin type interactions, and
 thus correct this shortcoming of our present approximation.

\section{Summary}

In this paper we have studied the NN interaction 
using a microscopic theory of quarks and diquarks.  
Nucleons were described as quark-diquark bound states. 
The quark and diquark degrees of freedom were integrated out using the 
path-integral method, and an effective Lagrangian was derived for 
mesons and baryons.  
The resulting trace-log formula contains various meson and 
nucleon interaction terms, including the $NN$ interaction
in the short range region expressed as a quark-diquark loop.  
Hence, the $NN$ interaction can be naturally expressed as meson 
exchanges at long ranges and quark or diquark 
exchanges at short ranges.  

For short range interactions, we have computed a quark-diquark loop 
corresponding to the direct term of the two-nucleon 
interaction.   
The resulting interaction is highly nonlocal.  
We have extracted scalar and vector type interactions 
in the local potential approximation.  
It turns out that the scalar term is attractive, while the vector 
is term repulsive.  

In the present paper our numerical calculations 
contained only scalar 
diquarks, as a first step toward a full calculation.  
Therefore, we have concentrated our study  mostly  
on the interaction ranges or, equivalently, the masses, 
because the magnitude of strengths is affected by 
the axial-vector diquark.  
On the other hand, the interaction ranges were better studied
in the present study including only scalar diquarks, reflecting the size 
of the nucleon.  
Consequently, the mass parameters of the interaction ranges 
for the scalar and vector interactions 
were found to be about 650 MeV for the scalar type  
and about 800 MeV for the vector type, once again 
with the nucleon size set at 0.77 fm.  
Hence in our model, the scalar-isoscalar interaction  
emerges from the quark-diquark loop at an energy scale similar to
that of the sigma and omega meson exchanges. 
The existence of the two components in the nuclear force, 
boson exchanges and quark-diquark exchanges, is a
general feature when we consider a model of nucleons 
that is composed of a quark (and diquark) core  
surrounded by meson cloud.

The present result encourages us to further study baryon
properties by extending 
the model to include the axial-vector diquark.  
We have already started such a study.  
When the axial-vector diquark is included, loop integrals diverge more
strongly than in the case of the scalar diquark, due to the massive
vector nature of the propagator.  
Although this causes the numerical study to be more complicated than
in the present case, 
this work is in progress.

\section*{Acknowledgments}
We would like to thank Veljko Dmitrasinovic for his careful reading
of this manuscript.
This work was supported in part by the Sasakawa Scientific Research Grant
from The Japan Science Society.


\begin{thebibliography}{99}


\bibitem{Lacombe:dr}
M.~Lacombe, B.~Loiseau, J.~M.~Richard, R.~Vinh Mau, J.~Cote, P.~Pires and R.~De Tourreil,
Phys.\ Rev.\ C {\bf 21}, (1980) 861.

\bibitem{Machleidt:hj}
R.~Machleidt, K.~Holinde and C.~Elster,
Phys.\ Rept.\  {\bf 149}, (1987) 1.


\bibitem{Toki:ai}
H.~Toki,
Z.\ Phys.\ A {\bf 294}, (1980) 173.

\bibitem{Oka:rj}
M.~Oka and K.~Yazaki,
Prog.\ Theor.\ Phys.\  {\bf 66}, (1981) 556; {\it ibid.} 572.

\bibitem{Takeuchi:yz}
S.~Takeuchi, K.~Shimizu and K.~Yazaki,
Nucl.\ Phys.\ A {\bf 504}, (1989) 777.


\bibitem{Fujiwara:2001pw}
Y.~Fujiwara, T.~Fujita, M.~Kohno, C.~Nakamoto and Y.~Suzuki,
Phys.\ Rev.\ C {\bf 65}, (2001) 014002.




\bibitem{Nagata:2003gg}
  K.~Nagata and A.~Hosaka,
  Prog.\ Theor.\ Phys.\  {\bf 111},  (2004) 857.

\bibitem{NJL}
Y. Nambu and G. Jona-Lasinio, Phys. Rev. {\bf 122}, (1961) 345; 
{\it ibid.} {\bf 124} (1961) 246.

\bibitem{Abu-Raddad:2002pw}
L.~J.~Abu-Raddad, A.~Hosaka, D.~Ebert and H.~Toki,
Phys.\ Rev.\ C {\bf 66}, (2002) 025206.  


\bibitem{Eguchi:1976iz}
T.~Eguchi,
Phys.\ Rev.\ D {\bf 14}, (1976) 2755.

\bibitem{Dhar:1983fr}
A.~Dhar and S.~R.~Wadia,
Phys.\ Rev.\ Lett.\  {\bf 52}, (1984) 959.

\bibitem{Ebert:1986kz}
D.~Ebert and H.~Reinhardt,
Nucl.\ Phys.\ B {\bf 271}, (1986) 188.

\bibitem{Ebert:1997hr}
D.~Ebert and T.~Jurke,
Phys.\ Rev.\ D {\bf 58}, (1998) 034001. 

\bibitem{Ishii:2000zy}
N.~Ishii,
Nucl.\ Phys.\ A {\bf 689}, (2001) 793

\bibitem{espriu}
D. Espriu, P. Pascual, and R. Tarrach, Nucl. Phys. B{\bf 214}, 285 (1983).



\bibitem{Hatsuda:1994pi}
T. Hatsuda and T. Kunihiro
Phys. Rept, {\bf 247}, 221 (1994).


\bibitem{Ebert:1994mf}
D. Ebert, H. Reinhardt and M. K. Volkov,
Prog. Part. Nucl. Phys, {\bf 33}, 1 (1994).

\bibitem{Vogl:1991qt}
U. Vogl and W. Weise,
Prog. Part. Nucl. Phys, {\bf 27}, 195 (1991).


\bibitem{Cahill:1987qr}
R. T. Cahill, C. D. Roberts and J. Praschifka,
Phys. Rev. {\bf D36}, 2804 (1987).

\bibitem{Cahill:1988zi}
  R.~T.~Cahill,
  Austral.\ J.\ Phys.\  {\bf 42}, 171 (1989).

\bibitem{Reinhardt:1989rw}
  H.~Reinhardt,
  Phys.\ Lett.\ B {\bf 244}, 316 (1990).

\bibitem{Ishii:rt}
N.~Ishii, W.~Bentz and K.~Yazaki,
Phys.\ Lett.\ B {\bf 318},  (1993) 26.



\bibitem{Nagata:2004ky}
  K.~Nagata, A.~Hosaka and L.~J.~Abu-Raddad,
  hep-ph/0408312.

\bibitem{Hess:1998sd}
  M.~Hess, F.~Karsch, E.~Laermann and I.~Wetzorke,
  Phys.\ Rev.\ D {\bf 58},  (1998) 111502.

\end{thebibliography}
\end{document}